\documentclass[12pt,twoside]{article}

\usepackage{epsf}
\usepackage{rotate}
\usepackage{cite}
\usepackage{times}
\usepackage{fancyheadings}
\advance \headheight by 3.0truept       

\renewcommand{\textfraction}{0}

\newlength{\nseparation}
\newenvironment{nfigure}
        {\begin{figure}[tb]\hrule\vspace{\nseparation}\par}
        {\vspace{\nseparation}\par \hrule \end{figure}}

\def\simge{\mathrel{\rlap{\raise 0.511ex \hbox{$>$}}{\lower 0.511ex \hbox{$\sim$}}}}
\def\simle{\mathrel{\rlap{\raise 0.511ex \hbox{$<$}}{\lower 0.511ex \hbox{$\sim$}}}} 
\def\slash#1{\setbox0=\hbox{$#1$}\dimen0=\wd0                     \setbox1=\hbox{/} \dimen1=\wd1 \ifdim\dimen0>\dimen1                    \rlap{\hbox to \dimen0{\hfil/\hfil}} #1                        \else                                       
      \rlap{\hbox to \dimen1{\hfil$#1$\hfil}}  
      /   \fi}                                         

\newcommand{\be}{\begin{equation}}
\newcommand{\ee}{\end{equation}}
\newcommand{\bea}{\begin{eqnarray}}
\newcommand{\eea}{\end{eqnarray}}
\newcommand{\msb}{\overline{\rm{MS}}}
\newcommand{\mev}{\ {\rm MeV}}   
\newcommand{\gev}{\ {\rm GeV}}   

\newcommand{\mud}{\overline{m}_{u,d}}
\newcommand{\ms}{\overline{m}_s}
\newcommand{\mb}{\overline{m}_b}
\newcommand{\epsp}{\varepsilon^{\prime}/\varepsilon}
\newcommand{\bbbar}{B\!-\!\bar{B}}
\newcommand{\kkbar}{K\!-\!\bar{K}}

\newcommand{\rhobar}{\overline{\rho}}
\newcommand{\etabar}{\overline{\eta}}

\begin{document}
\thispagestyle{empty}
\begin{flushright}
RM3-TH/00-15
\end{flushright}
\vskip 1.5cm

\begin{center}
\boldmath
\textbf{\Large
Phenomenology of the Standard Model from Lattice QCD\footnote{
Invited talk at the
\emph{XX Physics in Collision Conference}, 
June 29 - July 1 2000, Lisbon, Portugal}}
\unboldmath
\end{center}
\vskip1.3cm 

\begin{center}
{\large Vittorio Lubicz}\\[3mm]
Dip. di Fisica, Univ. di Roma Tre and INFN, Sezione di Roma Tre, \\
Via della Vasca Navale 84, I-00146 Rome, Italy \\
{\small E-mail: lubicz@fis.uniroma3.it}
\end{center}
\vskip1.cm

\begin{center}
\textbf{\large Abstract}
\end{center}
Some recent results of lattice QCD calculations which are relevant for the 
phenomenology of the Standard Model are reviewed. They concern the lattice 
determinations of quark masses, studies of $\kkbar$ and $\bbbar$ mixings, and a 
prediction of the $B_s^0$-mesons lifetime difference. The results of a recent 
analysis of the CKM unitarity triangle, which is mostly based on the lattice 
calculations of the relevant hadronic matrix elements, are also presented.

\section{Introduction}\label{sect:intro}
\renewcommand{\textfraction}{0}
An accurate determination of the Standard Model free parameters in the quark 
sector, namely quark masses and CKM mixing angles, is a task of fundamental 
importance facing both experimentalist and theoretical particle physicist. Among
these parameters, for instance, the charm and bottom quark masses enter through 
the heavy quark expansion the theoretical expressions of several cross sections
and decay rates. The elements of the CKM mixing matrix and the angles of the 
unitarity triangle control the intensity of hadron weak decays and the mixing 
amplitudes of $K$ and $B$ mesons. The area of this triangle defines the extent 
of CP-violation in the Standard Model. On the experimental side, a major advance
in this context has been represented by the recent determination of the 
parameter $\epsp$\cite{na48,ktev}, which controls direct CP-violation in $K\to
\pi\pi$ decays. This measurement provides evidence of a CKM origin of 
CP-violation, and the evaluation of $\epsp$ within the Standard Model, or some 
of its proposed extensions, still represents a challenging theoretical task. 
From a more theoretical point of view, an accurate determination of quark masses
and mixing angles may give insights on the physics of flavour, revealing 
relations between masses and mixing angles, or specific textures in the quark 
mass matrix, which, if there, should be the consequence of a still undiscovered 
flavour symmetry.

Because of the confining property of strong interactions, the determination of 
quark masses and mixing angles requires a non-perturbative control of hadron
dynamics. A major role, in this context, has been played by the numerical 
simulations of lattice QCD which have reached, in the last years, an accuracy 
unpaired by any other approach. With respect to other non-perturbative
techniques, like QCD sum rules or the $1/N$ expansion, lattice calculations 
allow a better control of the systematic errors, which may be (and has been) 
systematically improved in time. Space-time discretization errors, which are 
inherent to lattice QCD calculations, have been reduced by the introduction of 
improved versions of the lattice QCD action\cite{sw}-\cite{alpha}, and the 
increase of computer power has allowed, in most of the cases, an extrapolation 
of the lattice results to the physical continuum limit. Another potential source
of systematic error comes from the truncation of the perturbative expansion in
the calculation of lattice renormalization constants (or mixing coefficients). 
For many physical quantities, like the quark masses or the $B$-parameters of 
four-fermion operators, this error has been reduced to a negligible amount by 
the use of non-perturbative renormalization 
techniques\cite{bochicchio}-\cite{npm_sf}, which have proved to be a crucial 
ingredient in increasing the accuracy of the lattice determinations. Most 
likely, the largest source of uncertainty in lattice calculations is due, at 
present, by the use of the quenched approximation, derived by neglecting the 
effects of virtual quark loops. With the advent of the last generation of 
supercomputer, however, several unquenched calculations have been already 
performed, albeit with typically two flavours of dynamical quarks. In the last 
year, the first unquenched calculation of the $b$-quark mass has been 
performed\cite{mbunq}, which also employs a non-perturbative renormalization 
technique. Other important unquenched results concern the calculation of the 
$B$-meson pseudoscalar decay constant $f_B$\cite{bernard}, which is relevant for
the phenomenological studies of $\bbbar$ mixing.

In this talk some recent results of lattice QCD calculations which are relevant 
for the determination of the Standard Model fundamental parameters and for the 
phenomenology of particle physics are reviewed. These results concern the 
determination of the light ($u$, $d$ and $s$) and bottom quark masses, the 
calculation of the heavy mesons decay constants, $f_{D_{(s)}}$ and 
$f_{B_{(s)}}$, and the calculation of the $B_B$ and $B_K$ parameters which 
control the amplitude of $\bbbar$ and $\kkbar$ mixings respectively. 
The results of a recent analysis of the CKM matrix and the unitarity triangle,
which is based on the lattice calculations of the relevant hadronic matrix
elements, are also presented. Finally, I will discuss the lattice prediction for
the $B_s^0$-mesons lifetime difference which is found, within the Standard
Model, to be possibly accessible to the experimental observation. For this
review, some of the compilations of lattice results have been updated with new 
determinations which were not yet available at the time of the PIC20 
conference.\footnote{The transparencies of the talk at the PIC20 conference are 
available at the following URL site:\\ 
{\tt http://www.lip.pt/pic20/Vittorio.Lubicz/}}

\section{Quark Masses}\label{sect:qmass}
Quark masses are fundamental parameters of the Standard Model which cannot be
directly measured in the experiments because, unlike leptons, quarks are 
confined inside the hadrons. Being free parameters of the Standard Model 
lagrangian, quark masses cannot be computed on the basis of theoretical 
considerations only. Their values can be determined by comparing the result of 
a theoretical calculation of a given physical quantity, which depends on quark 
masses, with the corresponding experimental value. Typically, for instance, the 
pion and kaon masses are used to compute the values of the up-down and the 
strange quark masses, whereas the $b$-quark mass is determined by computing on 
the lattice the mass of the $B$ or the $\Upsilon$ mesons. Different choices are 
all equivalent in principle, and the differences in the values of quark masses, 
obtained by using different hadron masses as input parameters, give an estimate 
of the systematic error involved in the calculation.

As any other free parameter of the Standard Model lagrangian, quark masses can 
be defined as effective couplings and, as such, are both renormalization scheme 
and scale dependent. A scheme commonly adopted for quark masses is the $\msb$ 
scheme, with a renormalization scale chosen in the short-distance region to make
this quantity accessible to perturbative calculations. It is a common practice 
to quote the values of the light quark masses at the renormalization scale 
$\mu=2$ GeV, whereas the heavy quark masses are usually quoted at the scale of 
the quark mass itself, e.g. $\mb(\mb)$.

\subsubsection*{Light Quark Masses}
In the last few years a big effort has been devoted to compute on the lattice
the value of the strange quark mass. The results of the more recent calculations
obtained, in the quenched approximation, by using as experimental input the 
physical kaon mass, are shown in fig.\ref{fig:msave}.
\begin{nfigure}
\centerline{
\epsfxsize=20pc 
\epsfbox{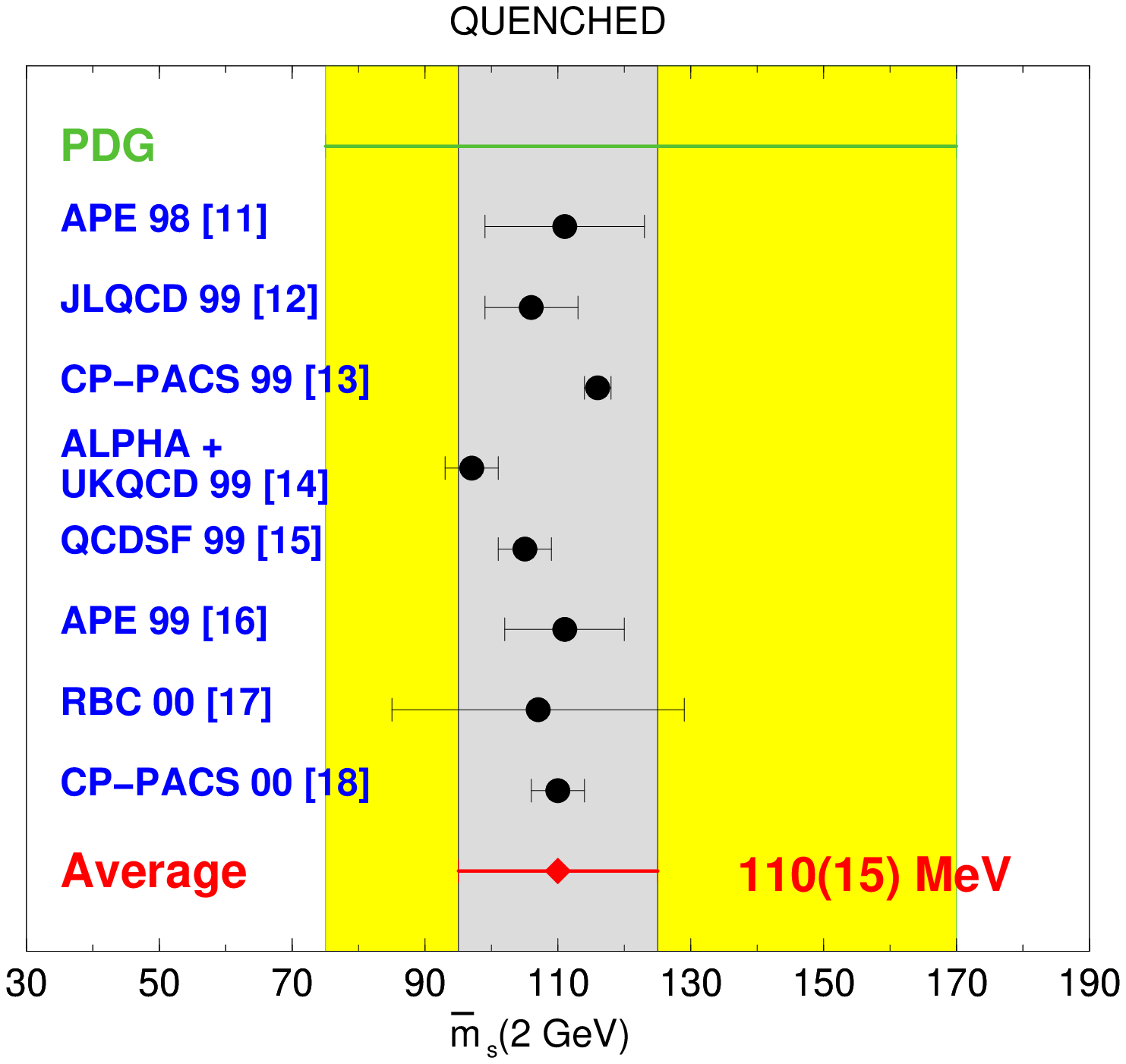}}
\caption{Values of the strange quark mass, $\ms(2\gev)$, obtained from recent 
lattice calculations in the quenched approximation (circles). The upper line is 
the estimate of $\ms$ quoted by the PDG\cite{pdg}, and the lowest point 
(diamond) represents the lattice (quenched) world average.}
\label{fig:msave}
\end{nfigure}
All these results have been obtained adopting a non-perturbative renormalization
technique, with the only exceptions of the two calculations by the CP-PACS 
collaboration\cite{mq_cppacs99,mq_cppacs00}, in which the quark mass 
renormalization constant has been evaluated by using one-loop perturbation 
theory. Previous experience suggests that an additional uncertainty of the order
of approximately 10\%, due to the use of perturbation theory, should be added to
the results quoted in refs.\cite{mq_cppacs99} and \cite{mq_cppacs00} to account
for the corresponding systematic error. Besides that, the results presented in 
fig.\ref{fig:msave} are also affected by other sources of systematics which are 
different among the several calculations, but their total effect may be 
estimated to be of the order of few per cent. For instance, the APE calculations
of refs.\cite{mq_ape98} and \cite{mq_ape99} do not involve the extrapolation to 
the continuum limit. In this case, the more extensive analysis of 
ref.\cite{mq_alpha99} suggests that the value of the strange quark mass is 
underestimated by approximately 3\%. Moreover, in the calculations of 
refs.\cite{mq_ape98} and \cite{mq_jlqcd99} the conversion from the 
non-perturbative RI-MOM renormalization scheme to the $\msb$ scheme has been 
performed by using N$^2$LO perturbation theory\cite{francolub}, since at the 
time when these studies have been performed the N$^3$LO result of 
ref.\cite{chetyrkin} was not available yet. In order to account for the 
difference between N$^2$LO and N$^3$LO, the results of refs.\cite{mq_ape98} and 
\cite{mq_jlqcd99} should be decreased by approximately 3\%. 

Within the quenched approximation, the main source of systematic error which
affects the determinations of the strange quark mass comes from the uncertainty
in fixing the physical lattice scale (i.e. the lattice spacing). Because of the
quenched approximation, different choices of the physical input, like the rho 
mass or the pion decay constant, lead to different estimates of the scale. This
introduces an additional uncertainty, which is of the order of 10\%, to the 
final estimate of the strange quark mass. By taking into account also this 
uncertainty, I quote as an average of the lattice results, within the quenched 
approximation,
\be
\ms(2\gev)^{\mbox{\scriptsize{\rm QUEN}}} = (110 \pm 15) \mev \, .
\label{eq:msq}
\ee 
This value is also shown in fig.\ref{fig:msave}, together with the lattice
results and the average value of $\ms$ quoted by the Particle Data 
Group\cite{pdg} (PDG). Notice that the uncertainty affecting the lattice 
determination of $\ms$ is approximately three times smaller than the one quoted 
by the PDG, although, in the former, the effect of the quenching approximation 
has not yet been taken into account. 

The average value of the up and down quark masses, $\mud \equiv\left(
\overline{m}_u + \overline{m}_d\right)/2$, can be also computed in a similar
way. It is convenient, however, to consider the ratio of the strange to the 
average up-down quark masses, because in this ratio many of the systematic 
uncertainties are expected to cancel out. On the lattice, this ratio is found 
to be in very good agreement with the value $24.4 \pm 1.5$ predicted by chiral 
perturbation theory\cite{leutwyler}. By using this information and 
eq.~(\ref{eq:msq}), I obtain:
\be
\mud(2\gev)^{\mbox{\scriptsize{\rm QUEN}}} = (4.5 \pm 0.6) \mev \, .
\label{eq:mudq}
\ee 

The remaining uncertainty in the determination of the light quark masses is the 
effect of quenching. At present, unquenched studies of light quark masses have
not yet reached the same degree of accuracy achieved in quenched calculations. 
In order to obtain an estimate of the quenching effect, it is thus convenient to
compute directly the ratio between the quenched and unquenched values of the 
quark mass, both obtained by using the same discretized version of the QCD 
action, the same renormalization procedure and the same choice of physical 
inputs to fix the quark mass itself and the lattice scale. To date, the most 
extensive unquenched calculation of the strange quark mass (with two flavours of
dynamical quarks) has been performed by CP-PACS\cite{mq_cppacs00}. They obtain 
the ratio $m^{\mbox{\scriptsize{\rm QUEN}}}_s/m^{\mbox{\scriptsize{\rm UNQ}}}_s
=1.25(8)$, suggesting a sizable decrease of the quark mass in the unquenched 
case. This result, however, is not confirmed by the other (although less 
accurate) unquenched studies, by SESAM\cite{mqu_sesam}, APE\cite{mqu_ape} and 
MILC\cite{mqu_milc}, which find a decrease in the unquenched case rather of the 
order of 10\%. Given the present situation, I believe that, in order to quote a 
final estimate of the lattice results, it is appropriate to include the
quenching error as an additional systematic uncertainty in eqs.~(\ref{eq:msq}) 
and (\ref{eq:mudq}), rather than varying the central values. Assuming this 
error to be of the order of 20 MeV in the case of the strange mass, I obtain: 
\be
\ms(2\gev) = (110 \pm 25) \mev \label{eq:ms}
\ee and
\be 
\mud(2\gev) = (4.5 \pm 1.0) \mev \label{eq:mud} \, .
\ee

\subsubsection*{The $b$-quark Mass}
Since the $b$-quark mass is larger than typical values of the ultraviolet
cutoff in present lattice calculations ($a^{-1} \sim 3$ GeV), the $b$-quark 
cannot be directly simulated on the lattice. However, the $b$ mass is also
larger than the typical energy scale of strong interactions, and the heavy 
degrees of freedom of the $b$-quark can be integrated out. Beauty hadrons may be
therefore simulated on the lattice in the framework of a low-energy effective 
theory. Indeed, in the past years, many lattice simulations of the Heavy Quark 
Effective Theory (HQET) and Non-Relativistic QCD (NRQCD) have been performed.

Within the effective theory, the mass of a $B$-meson, $M_B$, is related to the 
pole mass of the $b$-quark through the relation:
\be
M_B = m^{\mbox{\scriptsize{\rm pole}}}_b + \varepsilon - \delta m
\label{eq:mbpole}
\ee
where $\varepsilon$ is the so-called binding energy, which can be computed
by a numerical simulation in the effective theory, and $\delta m$ is the 
residual mass, generated, in the effective theory, by radiative corrections. The
perturbative calculation of the residual mass and the non-perturbative 
calculation of the binding energy thus allow a determination of the pole mass of
the $b$-quark. The result can be then translated into the $\msb$ mass, $\mb$, by
using again perturbation theory. 

An important observation, concerning this procedure, is that the binding energy
$\varepsilon$ is not a physical quantity, and it is affected indeed by a power
divergence proportional to the inverse lattice spacing, $1/a$. This divergence 
is canceled by a similar singularity in the residual mass $\delta m$. Moreover, 
the pole mass $m^{\mbox{\scriptsize{\rm pole}}}_b$, and consequently $\delta m$,
are also affected by a renormalon singularity, which introduces in their 
definitions an uncertainty of the order of $\Lambda_{QCD}$. This singularity is 
then canceled by the perturbative series relating the pole mass and the $\msb$ 
mass. As a result, the $\msb$ mass, $\mb$, is a finite, well defined, 
short-distance quantity. In the actual calculation, however, since the residual 
mass is computed up to a finite order in perturbation theory, only a partial 
cancellation of both the power divergence and the renormalon singularity may 
occur. For this reason, it is crucial in this calculation to compute the 
residual mass $\delta m$ up to the highest possible order in perturbation 
theory. 

At present, the most accurate determination of $\delta m$ has been obtained in
the framework of HQET. The two-loop analytical calculation has been performed in
ref.\cite{martisach}. In the quenched case, also the three-loop coefficient of 
the perturbative expansion has been evaluated\cite{direnzo}, by using a
numerical technique called numerical stochastic perturbation theory. The result 
is confirmed by the (less accurate) determination of ref.\cite{trottier}, 
obtained with a completely different approach, by fitting the results of small 
coupling Monte Carlo calculations. These combined theoretical efforts allow a 
determination of the $b$-quark mass which is accurate, in the quenched 
approximation, up to the N$^3$LO. Two independent results have been obtained so
far:
\bea
&& \mb^{\mbox{\scriptsize{\rm \ QUEN}}} = \left( 4.30 \pm 0.05 \pm 0.05 \right) 
\gev \qquad\qquad \left[\citen{mbn3lo_gim}\right] \nonumber \\
&& \mb^{\mbox{\scriptsize{\rm \ QUEN}}} = \left( 4.34 \pm 0.03 \pm 0.06 \right) 
\gev \qquad\qquad \left[\citen{mbn3lo_dav}\right]
\label{eq:mbquen}
\eea
nicely in agreement within each other. The last error in eqs.~(\ref{eq:mbquen})
represents the residual uncertainty due to the neglecting of higher orders in
perturbation theory. This uncertainty was estimated to be approximately 200 MeV 
and 100 MeV at the NLO\cite{mbnlo_gim} and N$^2$LO\cite{martisach} respectively.
By using NRQCD, results compatible with those in eqs.~(\ref{eq:mbquen}) have
been obtained. However, being only accurate at NLO, they are affected by a 
larger theoretical uncertainty.

The first unquenched calculation of the $b$-quark mass has been performed this 
year\cite{mbunq}. The result, which is accurate at the N$^2$LO, since the third 
coefficient of the perturbative expansion of $\delta m$ is yet unknown in the 
unquenched case, reads:
\be
\mb = \left( 4.26 \pm 0.06 \pm 0.07 \right) \gev \, .
\label{eq:mbunq}
\ee
This estimate represents to date the most accurate determination of the 
$b$-quark mass from lattice QCD calculations. Remarkably, the relative 
uncertainty is reduced at the level of 2\%.

\section{$\bbbar$ mixing, $\kkbar$ mixing and the Unitarity Triangle from 
Lattice QCD}\label{sect:ckm}
One of the most important test of the Standard Model, and a powerful tool for 
the search of new physics, is the analysis of the CKM unitarity triangle. At
present, this analysis is based on the study of four different constraints, 
coming from the experimental determinations of the following quantities: the 
ratio $|V_{ub}/V_{cb}|$, which determines the relative rate of $b \to u$ and 
$b \to c$ transitions in semileptonic decays, the neutral $B$-meson mass 
differences, $\Delta m_d$ and $\Delta m_s$, which control the frequencies of 
$B_d\!-\!\bar{B_d}$ and $B_s\!-\!\bar{B_s}$ oscillations, and the parameter 
$\varepsilon_K$, which defines the extent of indirect CP violation in kaon 
decays. Within the Standard Model, all these quantities are functions of the 
four Wolfenstein parameters of the CKM matrix, $A$, $\lambda$, $\rhobar$ and 
$\etabar$:
\bea
&& \left | \frac{V_{ub}}{V_{cb}} \right | = \frac{\lambda}{1-\lambda^2/2}
\sqrt{\rhobar^2+\etabar^2} \ , \label{eq:vubsuvcb} \\
&& \Delta m_d = C_B \ m_{B_d} \ f_{B_d}^2 \hat B_{B_d} \ A^2 \lambda^6 \ 
[(1-\rhobar)^2+\etabar^2] \ , \label{eq:deltam} \\
&& \frac{\Delta m_d}{\Delta m_s} = \frac{m_{B_d}f_{B_d}^2 \hat B_{B_d}}
{m_{B_s}f_{B_s}^2 \hat B_{B_s} }\ \lambda^2 \ [(1-\rhobar)^2+\etabar^2] \ ,
\label{eq:dms} \\
&& \left|\varepsilon_K \right| = C_K \ \hat B_K \ A^2 \lambda^6 \ \etabar 
\left[ A^2 \lambda ^4 \left(1-\rhobar\right) \ F_{tt} + F_{tc} \right] \ . 
\label{eq:epsk} 
\eea
\begin{nfigure}
\centerline{
\epsfxsize=26pc 
\epsfbox{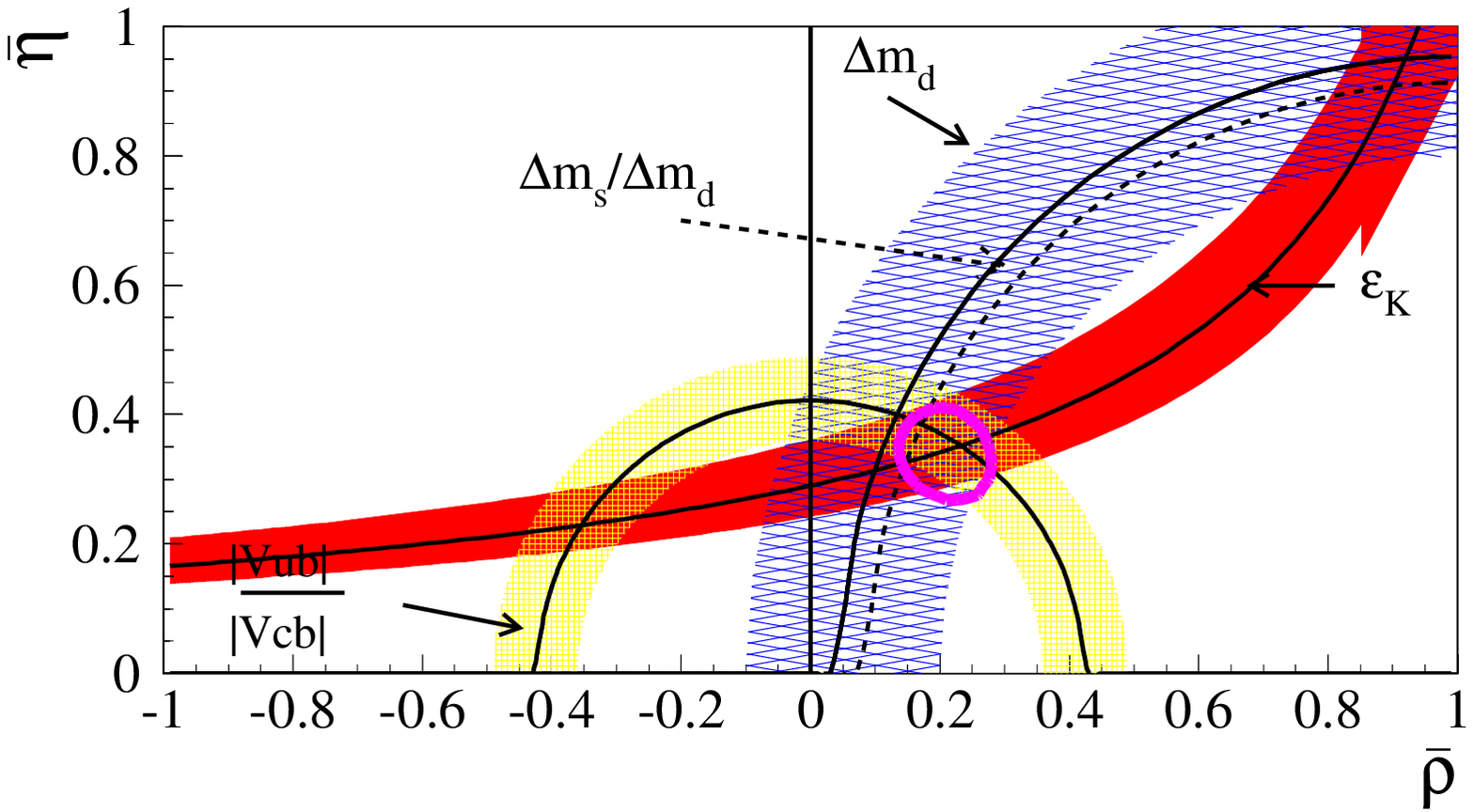}}
\caption{The allowed region for $\rhobar$ and $\etabar$ is shown together with
the uncertainty bands (at 68\% probability for $|V_{ub}/V_{cb}|$, 
$|\varepsilon_K|$, $\Delta m_d$ and the limit on $\Delta m_d/\Delta m_s$. The
picture is taken from ref.\cite{osaka}.}
\label{fig:bande}
\end{nfigure}
The coefficients $C_{B,K}$ and $F_{tt,tc}$ are known quantities, while
$f_{B_{d,s}}$, $\hat B_{B_{d,s}}$ and $\hat B_K$ are the pseudoscalar decay 
constants and $B$-parameters which encode, in the above equations, the 
non-perturbative effects of the strong interactions. For each quantity in 
eqs.~(\ref{eq:vubsuvcb})-(\ref{eq:epsk}), the comparison of the theoretical
expression with the corresponding experimental measurement defines a curve 
in the $\rhobar$-$\etabar$ plane which, in reality, because of the experimental 
and theoretical uncertainties, becomes a band in this plane. Consistency of the
Standard Model requires that the four bands, corresponding to the different
constraints, all intersect each other in the same region, which in turn defines
a set of allowed values for the $\rhobar$ and $\etabar$ parameters. A recent
analysis based on this approach has been performed in ref.\cite{osaka}, and the
results are illustrated in fig.~\ref{fig:bande}. Quantitative estimates of 
$\rhobar$ and $\etabar$, and the angles of the unitarity triangle, will be given
at the end of this section. 

The main theoretical issue in the analysis of the CKM triangle is the
non-perturbative calculation of the pseudoscalar decay constants and 
B-parameters entering in eqs.~(\ref{eq:vubsuvcb})-(\ref{eq:epsk}). Lattice QCD 
provides the optimal tool to perform such calculations and, in the following, I 
will review the most recent results of the lattice studies.

\subsubsection*{$\bbbar$ Mixing: Decay Constants and B-parameters}
The reliability of lattice calculations in computing the values of the 
pseudoscalar decay constants of $D$ and $B$ mesons has been recently supported 
by the experimental measurement of $f_{D_s}$, a quantity which has been computed
on the lattice since many years. In 1988, one of the first lattice calculation
of $f_{D_s}$ predicted, in the quenched approximation, the value $f_{D_s}=(215 
\pm 17)$ MeV\cite{gavela}. Subsequently, as shown in fig.~\ref{fig:fds}, the 
lattice predictions for $f_{D_s}$ have been always very stable in time. All 
lattice results obtained in the quenched approximation, in a period which 
extends over approximately 12 years, are presented in the figure as a function 
of the time, and the present experimental average is also shown for comparison.
\begin{nfigure}
\centerline{
\epsfxsize=20pc 
\epsfbox{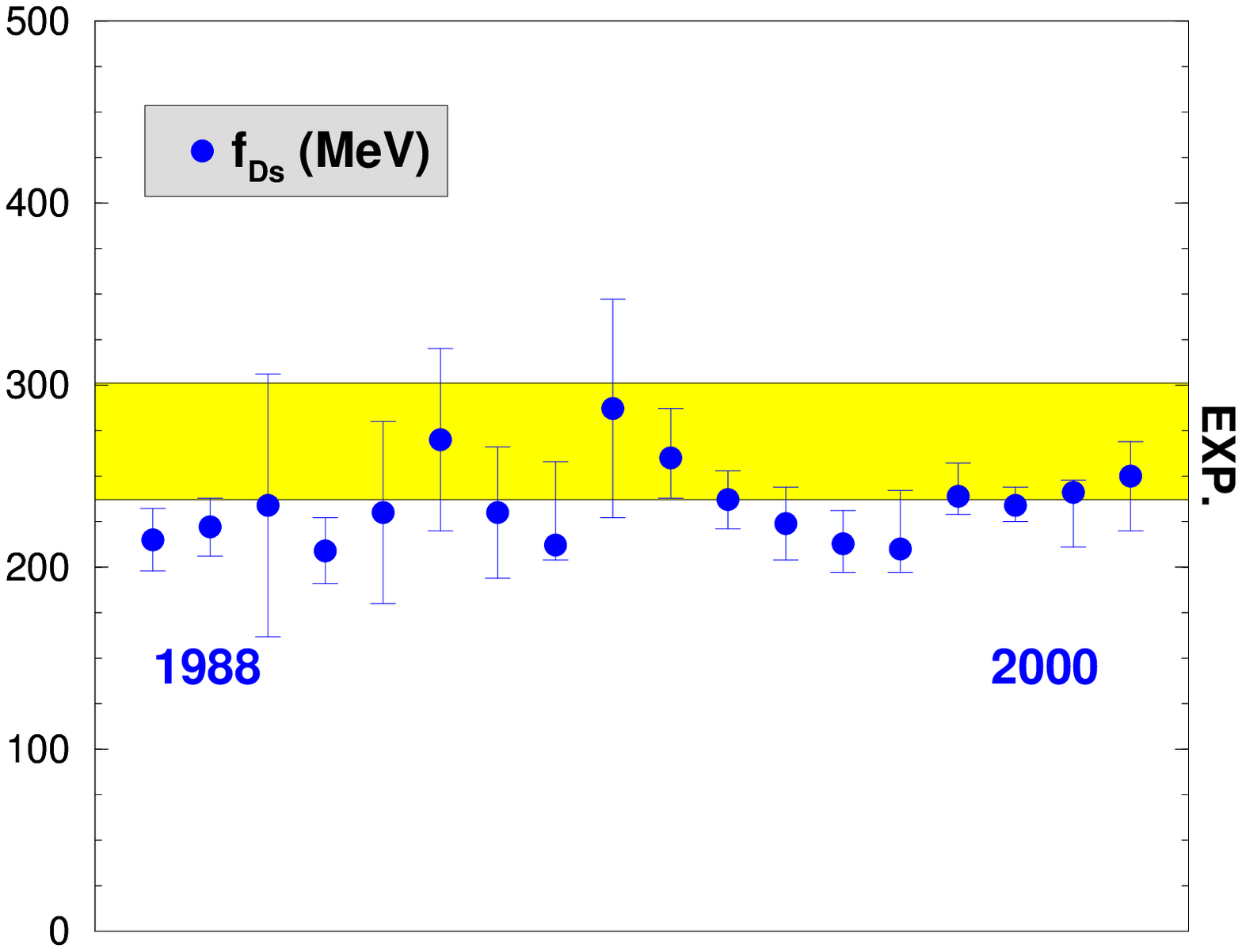}}
\caption{Quenched lattice results for the pseudoscalar decay constant $f_{D_s}$ 
plotted as a function of time. The band indicates the present experimental
average.}
\label{fig:fds}
\end{nfigure}
From the most recent lattice determinations of $f_{D_S}$, I obtain the quenched 
average:
\be
f^{\mbox{\scriptsize{\rm \ QUEN}}}_{D_s} = (235 \pm 20) \mev
\label{eq:fdsquen}
\ee
Unquenched calculations of $f_{D_S}$ have been performed by the 
MILC\cite{bernard} and CP-PACS\cite{fb_cppacs00} collaborations. They find an 
increase of the decay constant, in the unquenched case, of approximately 10\%, 
namely $f^{\mbox{\scriptsize{\rm UNQ}}}_{D_s}/f^{\mbox{\scriptsize{\rm QUEN}}}
_{D_s}=1.09(14)$ (MILC) \footnote{The error is my estimate based on the MILC 
results.} and $1.07(5)$ (CP-PACS). This correction can be then included in
eq.~(\ref{eq:fdsquen}) to obtain, as a final estimate of $f_{D_S}$ from lattice
calculations, the value:
\be
f_{D_s} = (250 \pm 25) \mev
\label{eq:fds}
\ee 
This prediction is in remarkable agreement with the present experimental 
average, $f_{D_s}=(271^{+30}_{-34})$ MeV\cite{fds_exp}. 

The calculation of the $B$-meson decay constant, $f_B$, which is not yet 
measured in the experiments, has been also the subject of intense activity of
lattice QCD simulations. The most recent results for $f_{B_d}$ and for the 
ratio $f_{B_s}/f_{B_d}$, obtained in the quenched 
approximation\cite{bernard,fb_cppacs00},\cite{fb_ape97}-\cite{fb_ukqcd00} 
are shown in fig.~\ref{fig:fbBb} (left).
\begin{nfigure}
\begin{tabular}{c c c} 
\epsfxsize8.0cm\epsffile{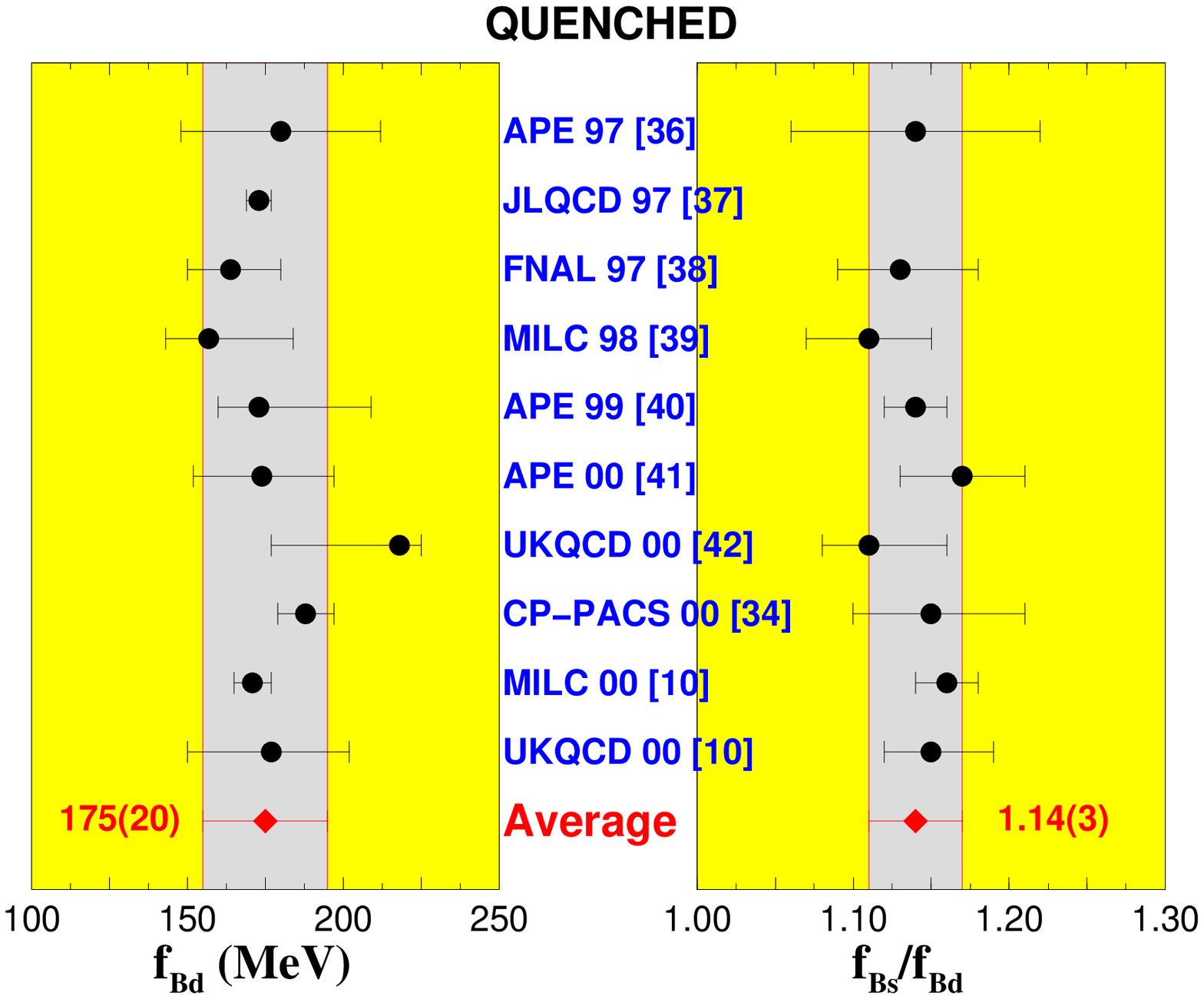} & \hspace*{-0.5cm} &
\epsfxsize8.0cm\epsffile{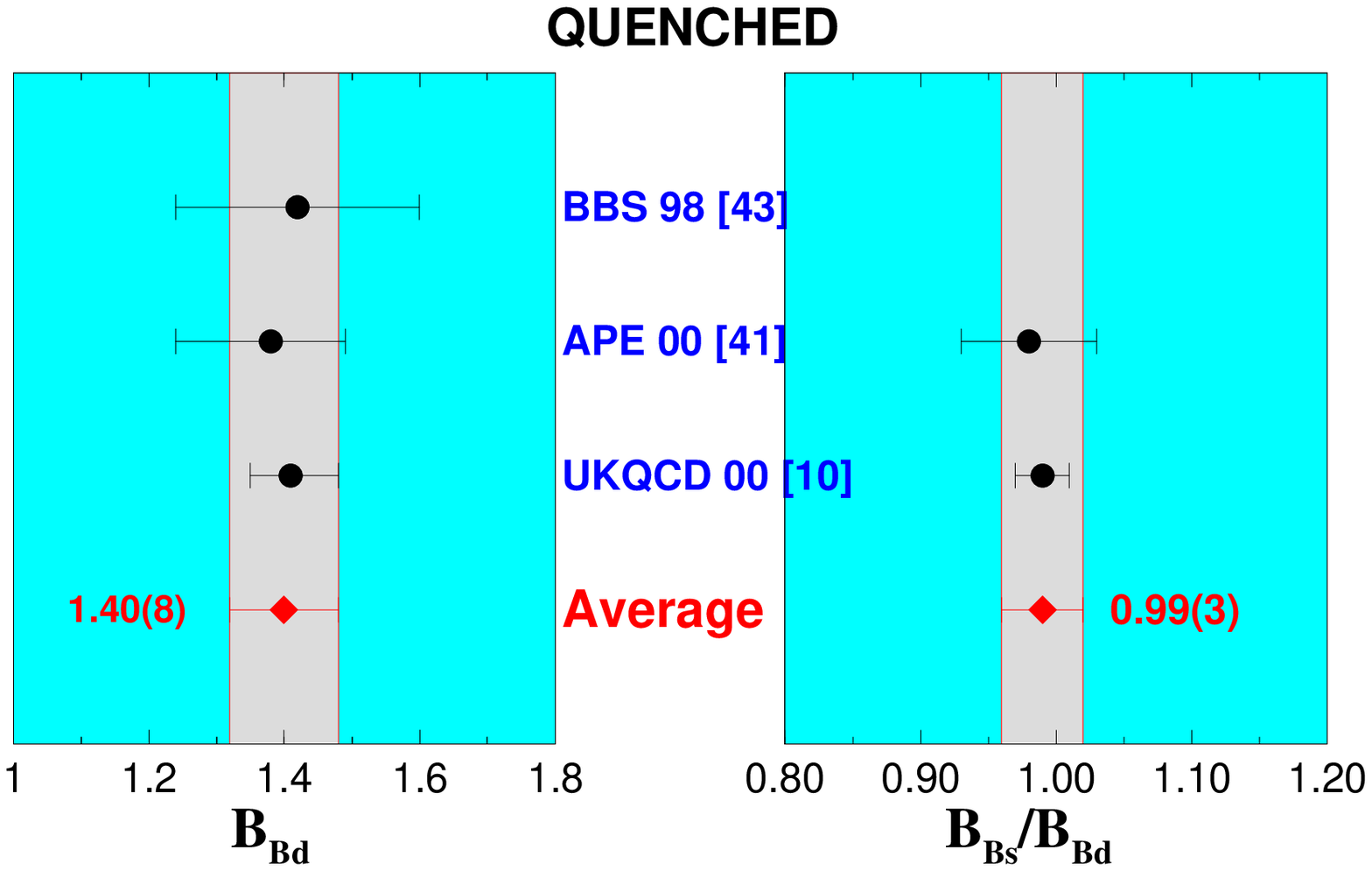}  \\
\end{tabular}
\caption{Values of $f_{B_d}$, $f_{B_s}/f_{B_d}$, $\hat B_{B_d}$ and 
$\hat B_{B_s}/ \hat B_{B_d}$ obtained, in the quenched approximation, from
recent lattice calculations.}
\label{fig:fbBb}
\end{nfigure}
Results for the decay constants have been also obtained by using 
NRQCD\cite{fbnqrcd_glock}-\cite{fbnqrcd_cppacs}. In this case, however, the 
predictions of different calculations are incompatible among each other, 
possibly signalling the presence of underestimated systematic effects. For this 
reason, the NRQCD results for the $B$-meson decay constants and for the 
$B$-parameters have not been included in the final lattice averages. From the 
results of fig.~\ref{fig:fbBb}, I obtain the estimates:
\be
f^{\mbox{\scriptsize{\rm \ QUEN}}}_{B_d} = (175 \pm 20) \mev \qquad , \qquad
(f_{B_s}/f_{B_d})^{\mbox{\scriptsize{\rm QUEN}}} = 1.14 \pm 0.03
\label{eq:fbquen}
\ee
which, for comparison, are also shown in the figure.

In the vacuum saturation approximation (VSA), the pseudoscalar decay constants
define the values of the matrix elements of the four-fermion operators which are
relevant for $\bbbar$ mixing. Deviations from the VSA are expressed, instead, by
the $B_B$ parameters. The recent quenched lattice determinations of $\hat 
B_{B_d}$ and $\hat B_{B_s}/ \hat B_{B_d}$ (the hat denoting the renormalization 
group invariant definition of these parameters) are shown in fig.~\ref{fig:fbBb}
(right). Although only few groups have performed such a calculation, the results
are in very good agreement among each other, so that one can derive the rather 
accurate averages shown in the figure, namely 
$\hat B^{\mbox{\scriptsize{\rm \ QUEN}}}_{B_d} = 1.40 \pm 0.08$ and 
$(\hat B_{B_s}/\hat B_{B_d})^{\mbox{\scriptsize{\rm QUEN}}} = 0.99 \pm 0.03$.

A source of uncertainty, in the lattice studies of $B$-physics, is introduced by
the necessity to extrapolate the results, reliably computed in the charm mass 
region, where discretization effects are under control, to the $b$-quark mass. 
In the case of the $B$-parameters, this uncertainty can be estimated, and 
partially reduced, by combining the relativistic results of fig.~\ref{fig:fbBb} 
with the prediction obtained in the infinite mass limit, $\hat 
B^{\mbox{\scriptsize{\rm \ HQET}}}_{B_d} = 1.29 \pm 0.08 \pm 0.06$\cite{reyes}. 
From this analysis, at the value of the $B$-meson mass, one gets $\hat B_{B_d}
\simeq 1.3\pm 0.1$. Since this estimate combines the results of different 
theories, affected by different systematic uncertainties, I prefer to quote, as 
a final central value for $\hat B_{B_d}$, in the quenched approximation, the 
average between the relativistic result and the one obtained combining with 
HQET, including in the systematic error the differences between the two 
determinations. As far as the ratio $\hat B_{B_s}/\hat B_{B_d}$ is concerned,
this is only marginally affected by the extrapolation, since $1/M$ corrections
are negligible in this case. One then obtains:
\be
\hat B^{\mbox{\scriptsize{\rm \ QUEN}}}_{B_d} = 1.36 \pm 0.10 \qquad , \qquad
(\hat B_{B_s}/\hat B_{B_d})^{\mbox{\scriptsize{\rm QUEN}}} = 0.99 \pm 0.03
\label{eq:Bbquen}
\ee

The above estimates should be improved to account for the effect of the quenched
approximation. Unquenched calculations of the pseudoscalar decay constants of 
$B$-mesons have been performed by the MILC\cite{bernard} and 
CP-PACS\cite{fb_cppacs00}collaborations. They find the ratio 
$f^{\mbox{\scriptsize{\rm UNQ}}}_{B_d}/f^{\mbox{\scriptsize{\rm QUEN}}}_{B_d}=
1.12^{+0.16}_{-0.11}$ (MILC) and $1.11 \pm 0.06$ (CP-PACS), in very good 
agreement within each other. For the ratio $f_{B_s}/f_{B_d}$, no significant 
difference between quenched and unquenched determinations has been observed by 
both MILC and CP-PACS, so that the quenching error on this quantity should be 
practically negligible. Starting from eq.~(\ref{eq:fbquen}), I then obtain, as 
final lattice estimates of the decay constants, the values:
\be
f_{B_d} = (200 \pm 25) \mev \qquad , \qquad
f_{B_s}/f_{B_d} = 1.14 \pm 0.03
\label{eq:fb}
\ee

Unfortunately, unquenched calculations of the $B_B$ parameters have not been 
performed to date. For these quantities, theoretical estimates based on quenched
chiral perturbation theory\cite{sharpe} suggest that the quenching error may be
of the order of 10\% at most. In the lack of a direct unquenched calculation,
I rely on these conservative estimates, and include this uncertainty in the 
systematic error, obtaining:
\be
\hat B_{B_d} = 1.36 \pm 0.17 \qquad , \qquad
\hat B_{B_s}/\hat B_{B_d} = 0.99 \pm 0.10
\label{eq:Bb}
\ee
Finally, combining eqs.~(\ref{eq:fb}) and (\ref{eq:Bb}), the lattice estimates 
for the two relevant parameters entering the theoretical expressions of $\Delta 
m_d$ and $\Delta m_d/\Delta m_s$ are derived:
\be
f_{B_d} \sqrt{\hat B_{B_d}} = (230 \pm 35) \mev
\label{eq:fbsqrtBb}
\ee
and
\be
\xi = \frac{f_{B_s} \sqrt{\hat B_{B_s}}}{f_{B_d} \sqrt{\hat B_{B_d}}} = 
1.14 \pm 0.06
\label{eq:csi}
\ee

In the standard analysis of the unitarity triangle, four constraints are used to
determine the values of only two parameters, $\rhobar$ and $\etabar$. Therefore,
an interesting possibility consists in relaxing, in turn, the theoretical 
estimate of one of the hadronic quantities relevant for the analysis, 
determining its value together with the values of the CKM parameters. In this 
way, a 68\% probability interval has been obtained for $f_{B_d} \sqrt{\hat 
B_{B_d}}$ in ref.\cite{osaka}:
\be
f_{B_d} \sqrt{\hat B_{B_d}} = (229 \pm 12) \mev
\label{eq:fit_fbsqrtBb}
\ee
which is nicely consistent with the lattice determination of 
eq.~(\ref{eq:fbsqrtBb}). Notice also that, at present, the analysis of the 
unitarity triangle, based on the lattice determinations of $\xi$ and $\hat B_K$,
provides an estimate of $f_{B_d} \sqrt{\hat B_{B_d}}$ which is more accurate 
than the direct theoretical determination. Therefore, a strong effort should be
put in lattice calculations to improve this estimate at the level of accuracy of
10\% or better.

\subsubsection*{$\kkbar$ Mixing and the $B_K$ parameter}
The $B_K$ parameter, which encodes the effects of strong interactions in the 
hadronic matrix element relevant for $\kkbar$ mixing, has been extensively 
studied in lattice QCD simulations. The two most accurate determinations, within
the quenched approximation, have been obtained by using staggered fermions, and 
they both involve an extrapolation to the continuum limit. The two results,
expressed in terms of the renormalization group invariant definition of the 
parameter, read:
\bea
&& \hat B^{\mbox{\scriptsize{\rm \ QUEN}}}_K = 0.86 \pm 0.04
\qquad\qquad \left[\citen{bk_gks}\right] \nonumber \\
&& \hat B^{\mbox{\scriptsize{\rm \ QUEN}}}_K = 0.87 \pm 0.06
\qquad\qquad \left[\citen{bk_jlqcd}\right]
\label{eq:bkquen}
\eea
in good agreement within each other. Several lattice calculations of $B_K$ have
been also performed by using Wilson fermions. The results are generally 
consistent with those in eqs.~(\ref{eq:bkquen}), but with larger systematic 
errors. This is due to the explicit breaking of chiral symmetry in the Wilson 
action which requires, in the definition of the renormalized operator, a 
delicate non-perturbative subtraction of the operators with different chirality.
A simple prescription to avoid the subtraction has been recently proposed in 
ref.\cite{bk_nosub}. If effective in practice, this procedure should allow to 
obtain, also with Wilson fermions, more accurate determinations of the kaon
$B$-parameter.

The quenching effect on $B_K$ has been estimated in ref.\cite{sharpe} by using 
quenched chiral perturbation theory, and a direct unquenched calculation has
been also performed in\cite{bk_kilcup}, at a fixed value of the lattice spacing.
Both studies estimate the quenching effect to be approximately 5\%. In order to 
take also into account the uncertainty on this estimate, the systematic error 
induced by the use of degenerate quark masses in the calculations of $\hat 
B_K$\cite{sharpe} and the discretization effects in the unquenched case, the 
total uncertainty is (conservatively) increased to 15\%. In this way, by using 
eqs.~(\ref{eq:bkquen}), one obtains: 
\be
\hat B_K = 0.87 \pm 0.14 \, .
\label{eq:bk}
\ee
As for the case of $\bbbar$ mixing, the lattice estimate of $\hat B_K$ is in
perfect agreement with the result of the overconstrained fit of the Standard
Model\cite{osaka}:
\be
\hat B_K = 0.89^{+0.21}_{-0.15} \, .
\label{eq:fit_bk}
\ee
This agreement provides additional evidence of the good level of accuracy 
reached, at present, by lattice calculations.

\subsubsection*{The Unitarity Triangle: a ``Lattice-based" Analysis}
The lattice determinations of $\hat B_K$, $f_{B_d} \sqrt{\hat B_{B_d}}$ and
$\xi$ provide the values of the input hadronic parameters entering the analysis 
of the unitarity triangle. One of these analysis, in which special attention has
been devoted to the determination of the related theoretical uncertainties, has 
been performed in ref.\cite{osaka}. The purpose of this study is to infer 
regions of the parameter space in which the values of the CKM parameters lie 
with given probabilities. In ref.\cite{osaka}, at 68\% probability, it is found
\be
\rhobar = 0.21 \pm 0.04  \ , \qquad \etabar = 0.34 \pm 0.04
\label{eq:rhoeta}
\ee
which imply, for the angles of the unitarity triangle, the values
\be
\sin(2\beta) = 0.72 \pm 0.07  \ , \qquad \sin(2\beta) = - 0.28 \pm 0.27 \ , 
\qquad \gamma = (59 \pm 7)^o \, .
\label{eq:angles}
\ee
The allowed region for the $\rhobar$ and $\etabar$ parameters is also shown in
fig.~\ref{fig:bande}. As can be observed from the figure, the Standard Model 
predictions are fully consistent with the experimental measurements of
semileptonic $b$-decays, $\bbbar$ mixing and $\kkbar$ mixing, within the present
level of theoretical and experimental accuracy. Moreover, the inferred value of
$\sin(2\beta)$ is consistent with the direct measurement from $J/\psi K_S$ 
events, by LEP, CDF, BaBar and Belle, which give the average $\sin(2\beta)=
0.52\pm 0.22$\cite{osaka}. Although the direct measurement has not yet reached 
a significant accuracy, important progresses are expected to come soon from the 
$B$-factories, thus allowing a crucial test of consistency with the unitarity 
triangle determination.

\section{The $B^0_s$-mesons Lifetime Difference}\label{sect:b0s}
In the Standard Model, the lifetime difference between the {\it short} and the 
{\it long} $B^0_s$-mesons is expected to be rather large, and possibly within 
reach for being measurable in the near future. In ref.\cite{dgbs_exp}, the 
following experimental bound has been obtained:
\be
\left(\frac{\Delta \Gamma_{B_s}}{\Gamma_{B_s}}\right)
^{\mbox{\scriptsize{\rm EXP}}} < 0.31 \qquad {\mathrm at \ 95\% \ CL}
\label{eq:dgbs_exp}
\ee
Theoretically, the prediction of $(\Delta \Gamma_{B_s}/\Gamma_{B_s})$ relies on 
the use of the operator product expansion, where the large scale is provided in
this context by the heavy quark mass. The theoretical expression for the width 
difference can be schematically written in the form:
\be 
\frac{\Delta \Gamma_{B_s}}{\Gamma_{B_s}} = K \, \left( G(z) - G_S(z) \, 
{\cal R} + \delta_{1/m} \right)
\label{eq:dgbs_th}
\ee
where $K$ is a known factor, and $G(z)$ and $G_S(z)$, with $z=m_c^2/m_b^2$, are 
Wilson coefficients which have been computed by including QCD radiative 
corrections at the NLO\cite{beneke1}. The factor $\delta_{1/m}$ contains the
subleading contribution in the $1/m_b$ expansion\cite{beneke2}. The 
non-perturbative strong interaction effects, in eq.~(\ref{eq:dgbs_th}), are 
encoded in the ratio of the matrix elements of two four fermion operators,
\be 
{\cal R} = \frac{\langle \bar B^0_s \vert Q_S \vert B^0_s \rangle}
{\langle \bar B^0_s \vert Q_L \vert B^0_s \rangle}
\label{eq:erre}
\ee
where $Q_L = \bar b \gamma_\mu ( 1 - \gamma_5) s \, \bar b \gamma_\mu ( 1 - 
\gamma_5) s$ and $Q_S = \bar b ( 1 - \gamma_5) s \, \bar b ( 1 - \gamma_5) s$.

At present, quenched lattice calculations of the ratio ${\cal R}$ have been 
performed in refs.\cite{bs_ape} and \cite{bs_hiro1}, the latter by using the 
NRQCD effective theory. The results of ref.\cite{bs_hiro1} have been 
subsequently corrected in ref.\cite{bs_hiro2}, and the two results are now 
nicely in agreement:
\bea
&& {\cal R} = - 0.93 \pm 0.03 ^{+0.00}_{-0.01}
\qquad\qquad\quad \left[\citen{bs_ape}\right] \nonumber \\
&& {\cal R} = - 0.91 \pm 0.03 \pm 0.12
\qquad\qquad \left[\citen{bs_hiro2}\right]
\label{eq:calR}
\eea
For the $B^0_s$-mesons lifetime difference, these measurements imply the 
prediction\cite{bs_ape}:
\be
\frac{\Delta \Gamma_{B_s}}{\Gamma_{B_s}} = ( 4.7 \pm 1.5 \pm 1.6) \times 
10^{-2} 
\label{eq:dgbs_lat}
\ee
which is small but possibly accessible to the experimental observation. It
should be noted that, despite the agreement between the results in
eq.~(\ref{eq:calR}), a much larger prediction for the width difference has been
obtained in ref.\cite{bs_hiro2}. The reason is that, in order to get a 
prediction for $\Delta \Gamma_{B_s}$ a quite large unquenched estimate of 
$f_{B_s}$ (namely $f_{B_s}=(245\pm 30)$ MeV, which is at the upper bound of the 
present lattice average) has been combined in ref.\cite{bs_hiro2} with the 
quenched determination (\ref{eq:calR}) of the matrix elements. As shown in 
ref.\cite{bs_ape}, however, a more accurate theoretical estimate of $\Delta 
\Gamma_{B_s}/\Gamma_{B_s}$ can be obtained in terms of the hadronic parameter 
$\xi$ which, compared to $f_{B_s}$, is much less affected by theoretical 
uncertainties. In this way, the prediction (\ref{eq:dgbs_lat}) has been derived.

\section{Conclusions}\label{sect:conclu}
The results of lattice QCD calculations are playing a crucial role in
determining the values of the Standard Model free parameters. The lattice 
determinations of quark masses have reached at present a high-level of 
statistical and systematical accuracy, which is of the order of 2\% in the case 
of the $b$-quark mass. For light quarks the accuracy is at the level of 10\%, 
although the effect of the quenching approximation still requires, in this case,
further investigations. The lattice studies of $\bbbar$ and $\kkbar$ mixings 
are of crucial importance for the determinations of the CKM matrix elements and 
for the analysis of the unitarity triangle. Moreover, the comparison between the
lattice predictions and the overconstrained fits of Standard Model, whenever 
available ($f_B\sqrt{B_B}$ and $B_K$), reveals an high degree of consistency. 

\section*{Acknowledgments}
I wish to thank Prof. Gaspar Barreira for the organization of this beautiful 
conference. I am grateful to my collaborators D.~Becirevic, V.~Gimenez and
G.~Martinelli for many interesting discussions on the topics covered by this
review. I acknowledges the M.U.R.S.T. and the INFN for partial support.

\end{document}